# Weighted Fuzzy-Based PSNR for Watermarking

Maedeh Jamali, Nader Karimi, Shadrokh Samavi
Department of Electrical and Computer Engineering,
Isfahan University of Technology,
Isfahan, 84156-83111, Iran

***Abstract*— One of the problems of conventional visual quality evaluation criteria such as PSNR and MSE is the lack of appropriate standards based on the human visual system (HVS). They are calculated based on the difference of the corresponding pixels in the original and manipulated image. Hence, they practically do not provide a correct understanding of the image quality. Watermarking is an image processing application in which the image's visual quality is an essential criterion for its evaluation. Watermarking requires a criterion based on the HVS that provides more accurate values than conventional measures such as PSNR. This paper proposes a weighted fuzzy-based criterion that tries to find essential parts of an image based on the HVS. Then these parts will have larger weights in computing the final value of PSNR. We compare our results against standard PSNR, and our experiments show considerable consequences.**

***Keywords—HVS, watermark, imperceptibility, weighted PSNR, fuzzy***

## I. INTRODUCTION

Due to the rapid expansion of information technology, digital content such as digital images has become familiar with our daily lives. They usually pass through some processing stages, and their quality is degraded before reaching the end-users. These end-users are often human observers. To maintain and improve the quality of images, assessing the quality is an essential point that should be considered in each stage.

Image quality assessment (IQA) is considered as a characteristic feature of an image, and the degradation of images is estimated by it. Usually, this degradation is measured compared to a reference image [1]. Practically, there are two ways for image quality evaluation: subjective and objective. As the end-users are human observers, the essential idea to evaluate images' perceptual quality is to score these distorted images (subjective) [2]. However, such subjective evaluation is typically time-consuming and costly, making them infeasible in image processing systems. Hence, IQA research aims to provide a simple objective evaluation of image quality that mimics the subjective human judgment [3].

Recently, IQA methods have been considered in many image processing applications. Full-reference evaluation methods such as peak signal-to-noise ratio (PSNR) and mean-square error (MSE) have, in some cases, failed to conform to human user opinion [4-5]. Many research works have been done in this field, the first of which is SSIM [6]. In SSIM, natural images are assumed to have a structure, and this criterion examines the structural differences between overlapping blocks using the three criteria of brightness, contrast, and structural correlation. Since this criterion decides on a combination of several features and blocks, it performs better than PSNR. It conveys a better evaluation of the images' visual quality to the viewer, but it can still be far from the human evaluation.

Watermark, as an image processing application, is evaluated based on the quality of the final image. Watermarking requires a criterion based on the HVS that provides more accurate values than conventional criteria such as PSNR. Sometimes we can see the output of a watermarking method has low PSNR but based on HVS, the result is acceptable and vice versa.

Something important in watermarking is to embed the data in part of the image with the lowest effect on imperceptibility and suitable robustness against attacks. In this paper, our concentration is on assessing the first requirement; visual quality evaluation. We need criteria that satisfy the mentioned need. We need to find the most important and relevant parts of images viewed by human eyes for this aim. We combine three features, saliency, edge concentration, and intensity, to find the sensitive part that human eyes can detect its degradation. A fuzzy system is used to produce a fuzzy map based on their importance. The fuzzy system models the uncertainty and inference similar to a human observer to help us have an importance map near human opinion. This fuzzy map indicates a weighted map of the image used to make a weighted PSNR.

The rest of the paper is organized as follows. Our proposed method and details are explained in section II. Section III contains our experimental results, and a comparison with normal PSNR is illustrated there. Eventually, the paper is concluded in section IV.

## II. PROPOSED METHOD

Our proposed method is explained in this section. We try to introduce a weighted PSNR suitable for watermark application and consider the human visual requirements in evaluating the watermarked image. Based on the HVS, the most crucial parts of the image can be regions that absorb the human eye's attention in an image. This feature can be satisfied in the salient part of the image. Therefore, we consider saliency detection as one part of our method to detect an image's essential area. Also, we know part of the image with more edges can hide the human eyes' changes and be the right place for robust watermark embedding. However, these parts should have lower weights in our final map to evaluate the image's visual quality. Intensity is another feature we used for making the weighting map. Because we know the high-intensity part of the image is a good place for embedding and human eyes are not applicable for changing detection. We introduce three features, saliency, edge concentration, and intensity, to make a weighted map for evaluation. In the following, each feature and its combination will be explained. Fig. 1 shows the block diagram of the proposed method.

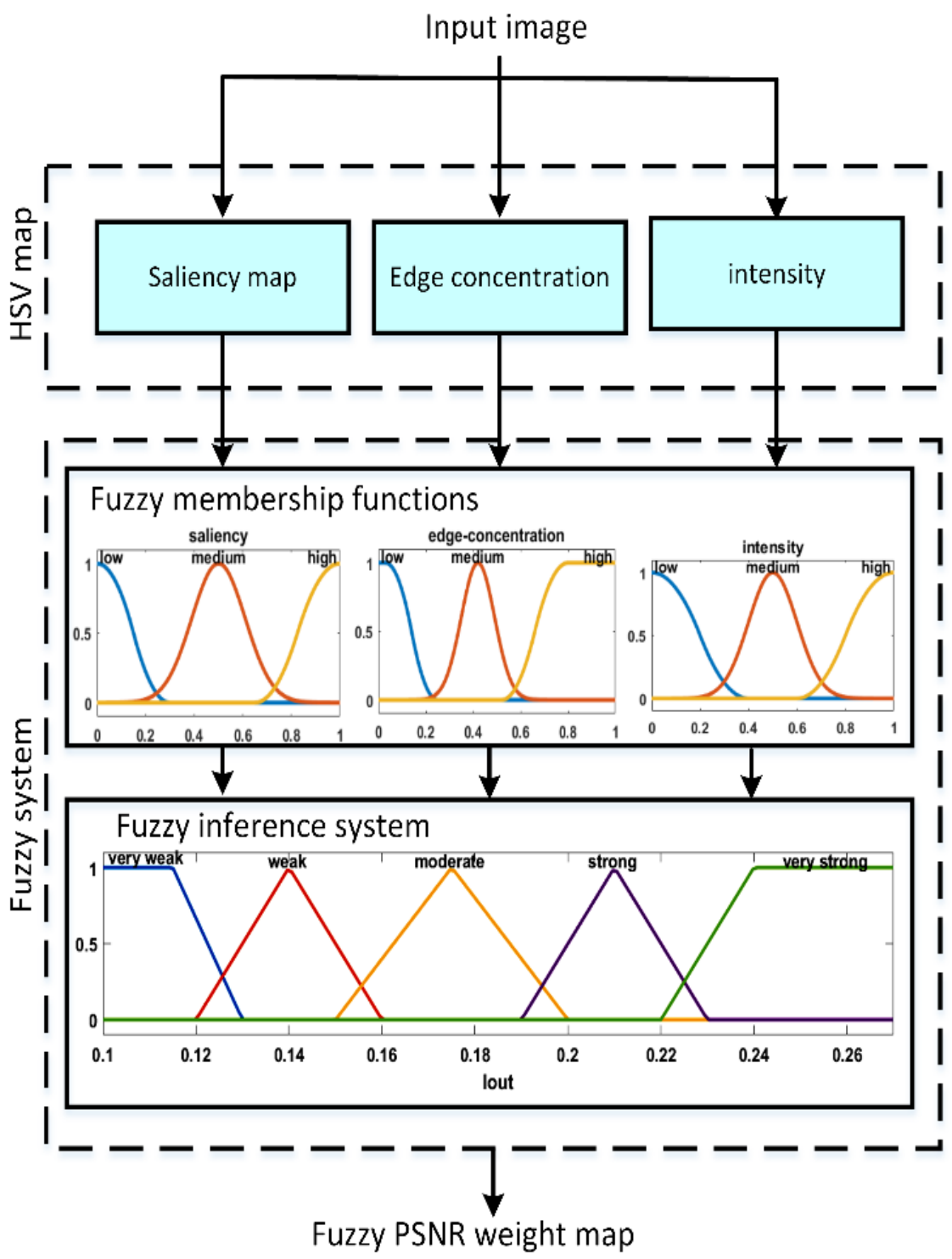

Figure 1. Block diagram of the proposed method

### A. *Saliency*

The basic idea in saliency detection is to find the most visually informative and important areas in an image. Visual saliency is an essential feature of HVS that tries to anticipate the most related regions of images viewed by human eyes. To show saliency methods can extract parts of the image near the HVS, we use several saliency detection methods [7-9], and their output is indicated in Fig.2. These methods do not need learning and are fast. We also asked some human observers to identify important areas in these images. Column (e) of Fig.2 shows the community of these areas, and the brighter areas are the areas that most people have selected. As we can see, these parts of the images that are of interest to the viewer are near the saliency algorithm's output. We use a weighted combination of [7] and [8] outputs to make the saliency map for feeding into a fuzzy system.

### B. *Edge concentration*

The human visual system is usually unable to detect changes in irregular areas of the image. Therefore, these regions are suitable for powerful embedding and should have a low weight in the final PSNR map. We select areas with a high concentration of edges as candidate regions. For this aim, we use the canny algorithm to find edges in an image. The canny output is divided into $8 \times 8$ blocks. For each point in a block, a $3 \times 3$ neighborhood is considered. Then the variance of edge pixels in this neighboring is calculated. The average value of all variance values is assigned to each block. Finally, we normalize the value to [0,1]. Fig. 3 shows the output of edge concentration.

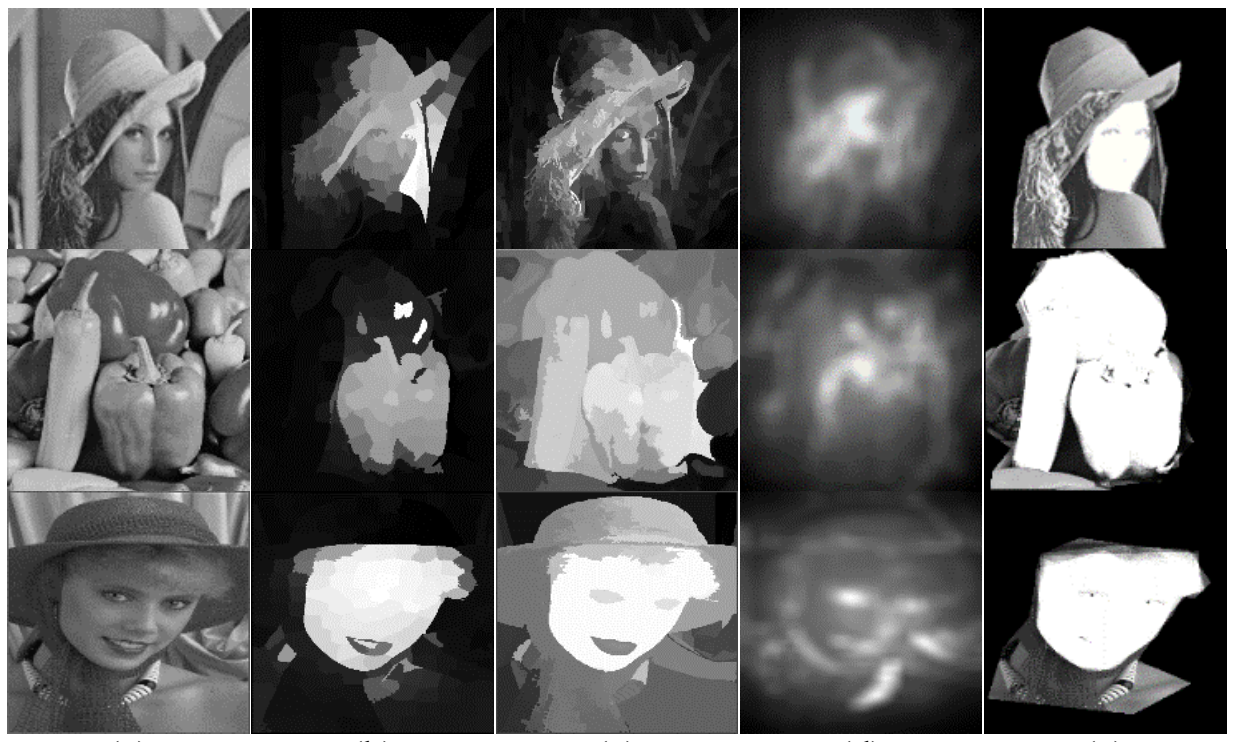

(a) (b) (c) (d) (e)

Figure 2. Comparing the output of different saliency detection algorithms with aggregating the views of the different human observer about important parts of the image, (a) Original image, (b) Output of [7], (c) Output of [8], (d) Output of [9], (e) Aggregation of human observers' output.

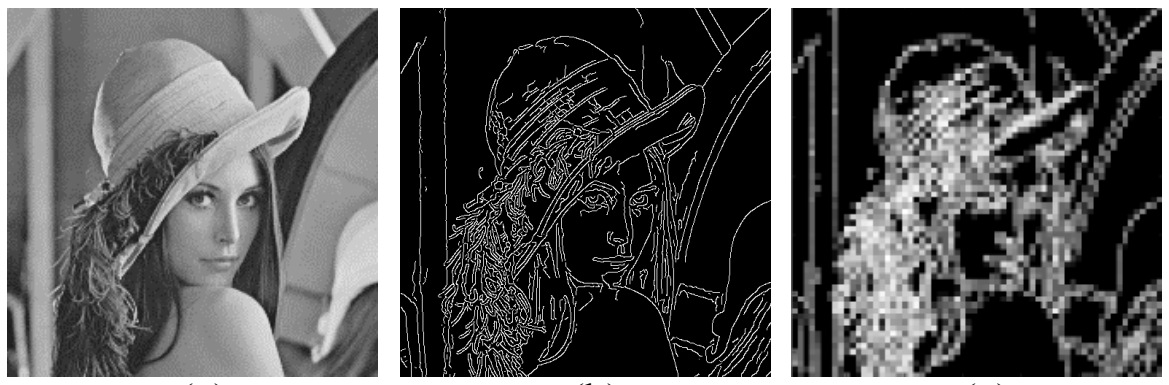

(a) (b) (c)

Figure 3. Edge-concentration computation, (a) original image, (b) canny output map, (c) normalized edge-concentration map.

### C. *Intensity*

Changes in regions of the image with higher intensity have a lower effect on human eyes, and our eyes practically cannot recognize them. This fact confirms the weber ratio [10] with this formula, $k = \Delta I \,/\, I$, that $I$ am the intensity. Because this ratio shows a constant value, so regions with higher intensity in an image should have considerable $\Delta I$ regard to lower intensity regions to be recognized. This means regions with higher intensity have lower weights in our final map because human eyes cannot distinguish these regions' changes. To find regions with high intensity, we divide the image into 8×8 blocks, and the average value of each block is considered the intensity of that block. Then we normalize the value into [0,1].

### D. *Fuzzy map*

After finding suitable features, we need a probability map that combines them. For this aim, we use a fuzzy system. Some features can conflict with each other. For example, an area can be salient while it has many edges, so probably the human visual system cannot recognize changes in that area. We can consider these situations in fuzzy systems. By defining different rules, all probable cases can be covered based on human expert opinion to have a map near the HVS [11].

We use Mamdani min-max fuzzy inference [12]. It is a simple method and suitable for systems that the rules are based on human knowledge, like our system.

### 1) *Fuzzy membership function*

We consider three parts in an image, high, medium, and low, that show the importance of regions based on the human observer. So, we need three fuzzy values related to these regions, and we will need three fuzzy membership functions for each mentioned feature. The membership functions have the task of mapping the input values to [0, 1] and in fact, determine how much the input value is correct [11]. In most cases, the choice of membership function depends on the type of issue and is determined in a heuristically, subjective, or objective way.

We use subjective membership function for intensity and saliency features. Human observers select three parts in an image after watermarking based on their importance rate (high, average, and low) for these membership functions. For the saliency membership function, we use the saliency map, and for intensity, the SSIM map is selected for human selection. Then probability functions based on their intensity will be calculated. Fig. 4 shows the membership function for saliency.

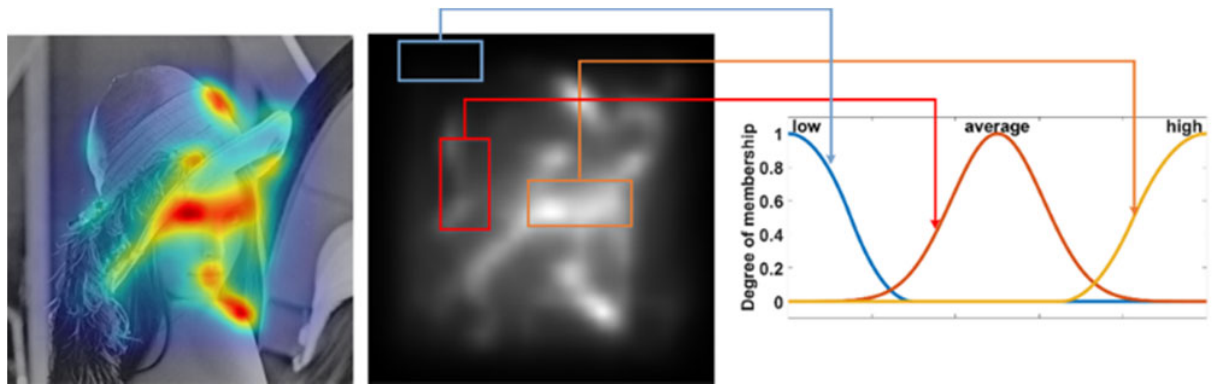

(a) (b) (c)

Figure 4. Saliency membership function, (a) saliency heat map, (b) saliency map, (c) final membership function.

We use Fuzzy C-Means (FCM) [13] for edge concentration membership function and is done based on data distribution. Due to the unknown data distribution of edge concentration date, first, we use FCM and experimentally cluster the data into nine centers. Then the probability distribution is fit into these clusters. However, we need three probability density function (PDF) at last. So, we try to merge these overlapped distributions and consider the data balance in each of them. Finally, the PDF will be scaled to become a normalized membership function with a maximum value of one. Fig. 5 shows the process of finding the membership functions for edge concentration.

We will have 27 rules based on the value of each feature and its combination. We consider five values for the fuzzy map that would divide the image into five regions based on their importance. Fig. 6 shows the final fuzzy output map. All the values in the fuzzy map are defuzzified into the range [0.1, 0.27]. This map weight all images based on their importance, and more important regions should be changed less than other regions. Hence, if we subtract the image's max value from all image values, we will make the final map to make weighted PSNR.

## E. *Weighted PSNR*

In visual evaluation, the goal is to assign more weight to the image's parts that are more important from a visual point of view to provide fewer values by evaluation criteria if these areas are manipulated. Because we use MSE in calculating PSNR, so we need to compute weighted MSE at first. Equation 1 shows the weighted MSE as follows:

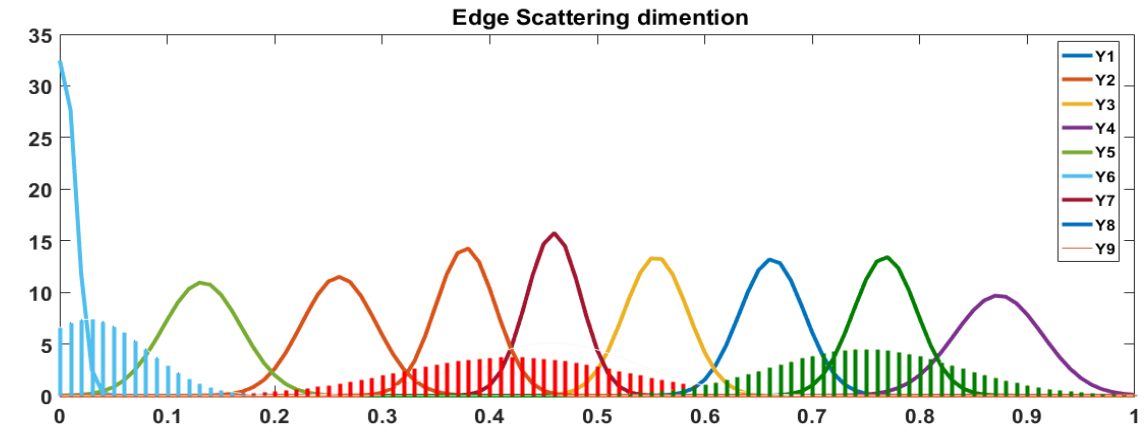

(a)

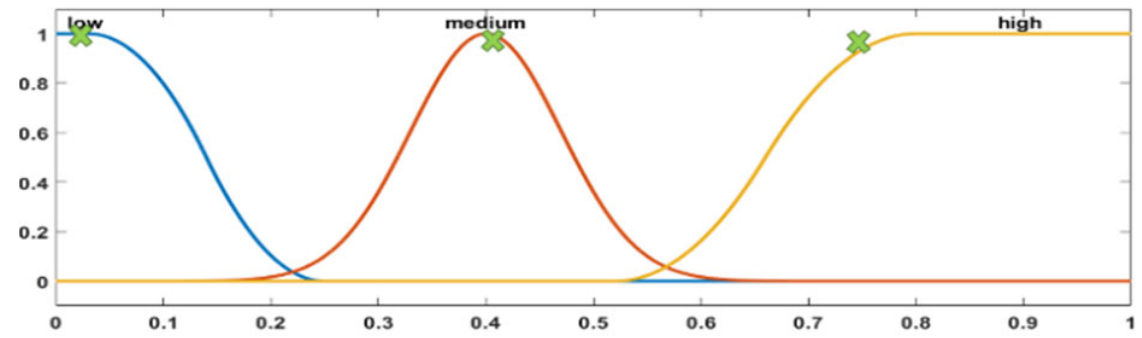

(b)

Figure 5. Edge concentration membership function, (a) FCM output, (b) normalized membership function

$$FMSE(x, y, FMap) = \frac{1}{N}\sum_{i=1}^{N}((x_i - y_i)^2.FMap) \quad (1)$$

where $FMap$ is the output of the fuzzy system, $x, y$ are the original and manipulated image, respectively, and $N$ is the number of pixels. After computing the weighted MSE, weighted fuzzy PSNR (WFPSNR) can be calculated as follows:

$$FPSNR(X, Y, FMap) = 10log_{10}(\frac{L^2}{FMSE(x, y)}) \quad (2)$$

where $L$ is the maximum value in the image that can be 255 or 1 based on the type of image.

## III. Experimental Results

As we mentioned before, we need to design HVS-based criteria for watermarking to evaluate our watermarked image's output from a visual point of view. So, we calculate WFPSNR and evaluate it against the watermarked image. We used a simple watermark algorithm. We divide the image into blocks with a size of 16×16. Then these blocks are DCT transformed. We select two random coefficients of some blocks in two different regions of Lena and change them. We select a part of Lena's face and a small area in the fur part of Lena's hat. The first part has a high degree of importance based on human observer visually, and the second part is a noisy part that changes are not practically detectable.

As we can see in Fig. 7, both images have approximately similar PSNR while these images have a different quality from HVS's point of view, and Fig. 7 (a), in comparison to Fig. 7 (b) has better visual quality. In Figure 7 (b), the user expects a

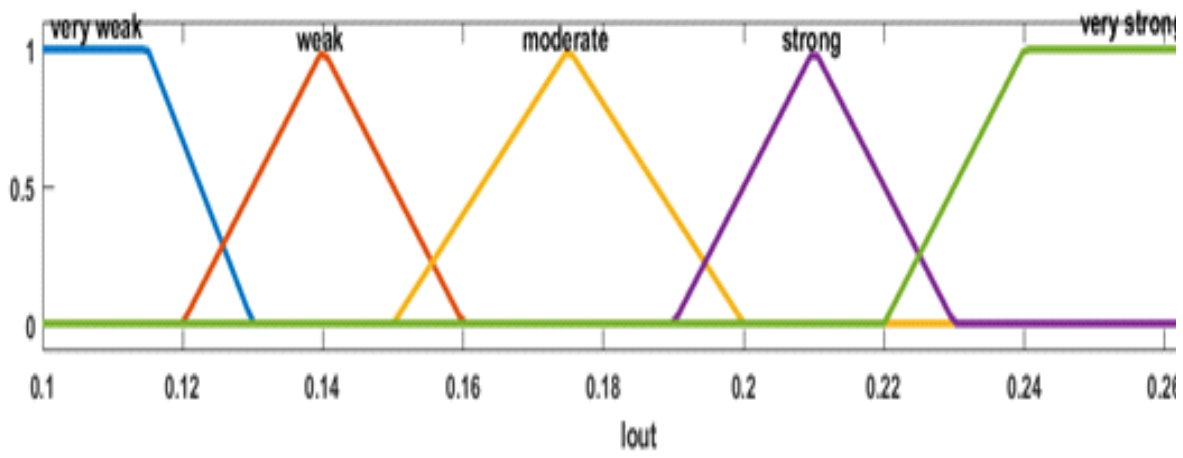

Figure 6. Fuzzy output map

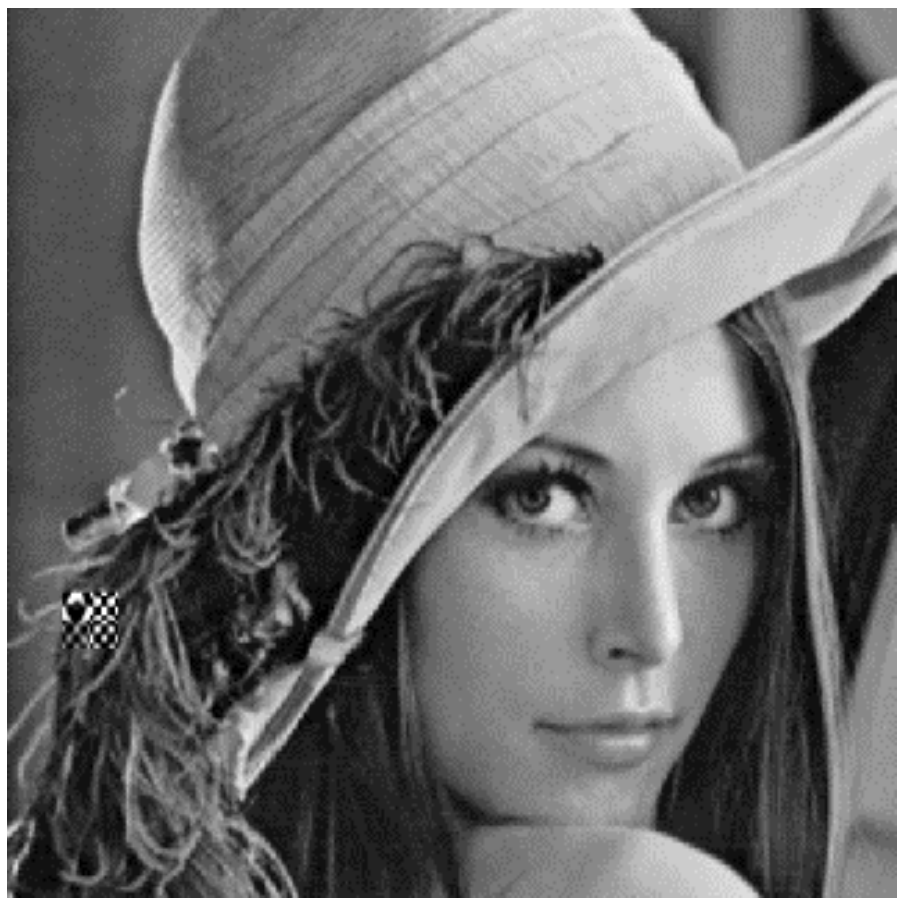

**WFPSNR=35.04**
PSNR=22.95
(a)

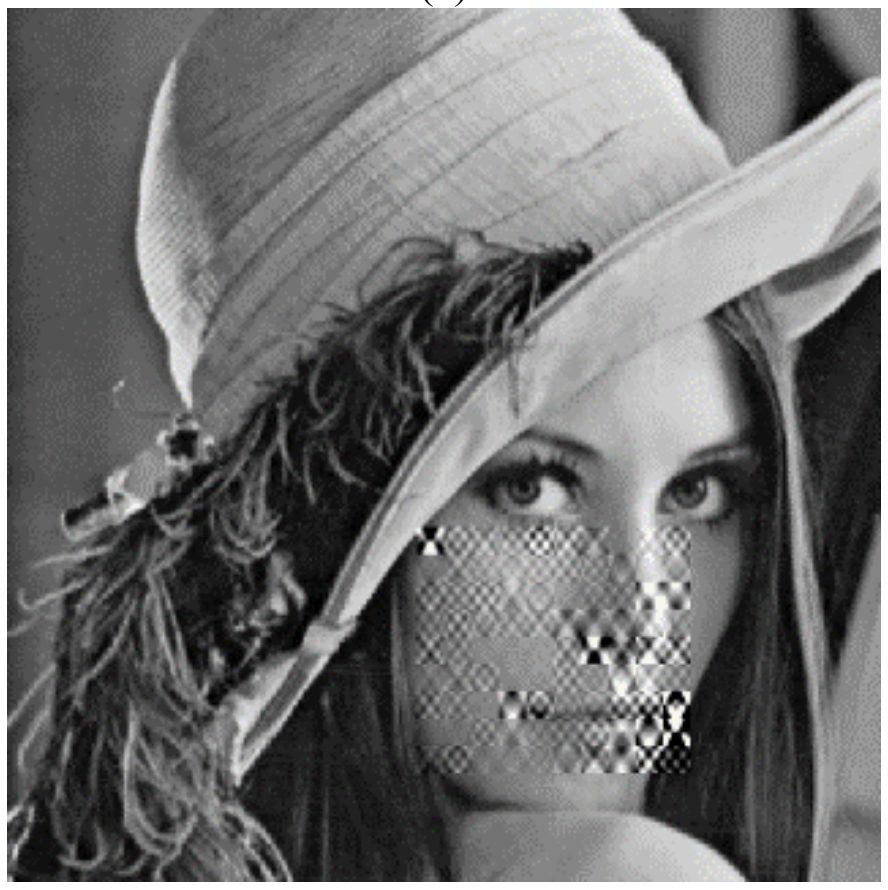

**WFPSNR=6.66**
PSNR=23.51
(b)

Figure 7. Comparison of standard PSNR and WFPSNR, (a) HSV based watermark algorithm, (b) simple watermark algorithm

lower PSNR due to the degradation of the face, but the PSNR values are near the PSNR value for Fig.7 (a).

However, we can see WFPSNR differentiate these images, and the reported values are based on HVS. For Fig.7 (a), because the important parts of the image are intact, so WFPSNR calculates a higher value. In Fig. 7 (b) WPSNR has a considerably lower value due to manipulations that have occurred to vital parts of the image. To better evaluate our proposed method, we consider some standard images in the watermark and applied image processing attacks such as salt-pepper (SP) and Gaussian noise (GN) on these images. In the first phase, we degraded the quality of images by applying these attacks on the important part of them, and in the next phase, the non-important part of these images from the visual point of view is attacked. Figure 8. Shows images and their PSNR values. When an essential part of an image is manipulated such that human eyes can detect them, the WFPSNR should indicate lower values than standard PSNR. For the next scenario, WFPSNR should have higher values. So, we can see the proposed weighted PSNR can satisfy this requirement based on HVS.

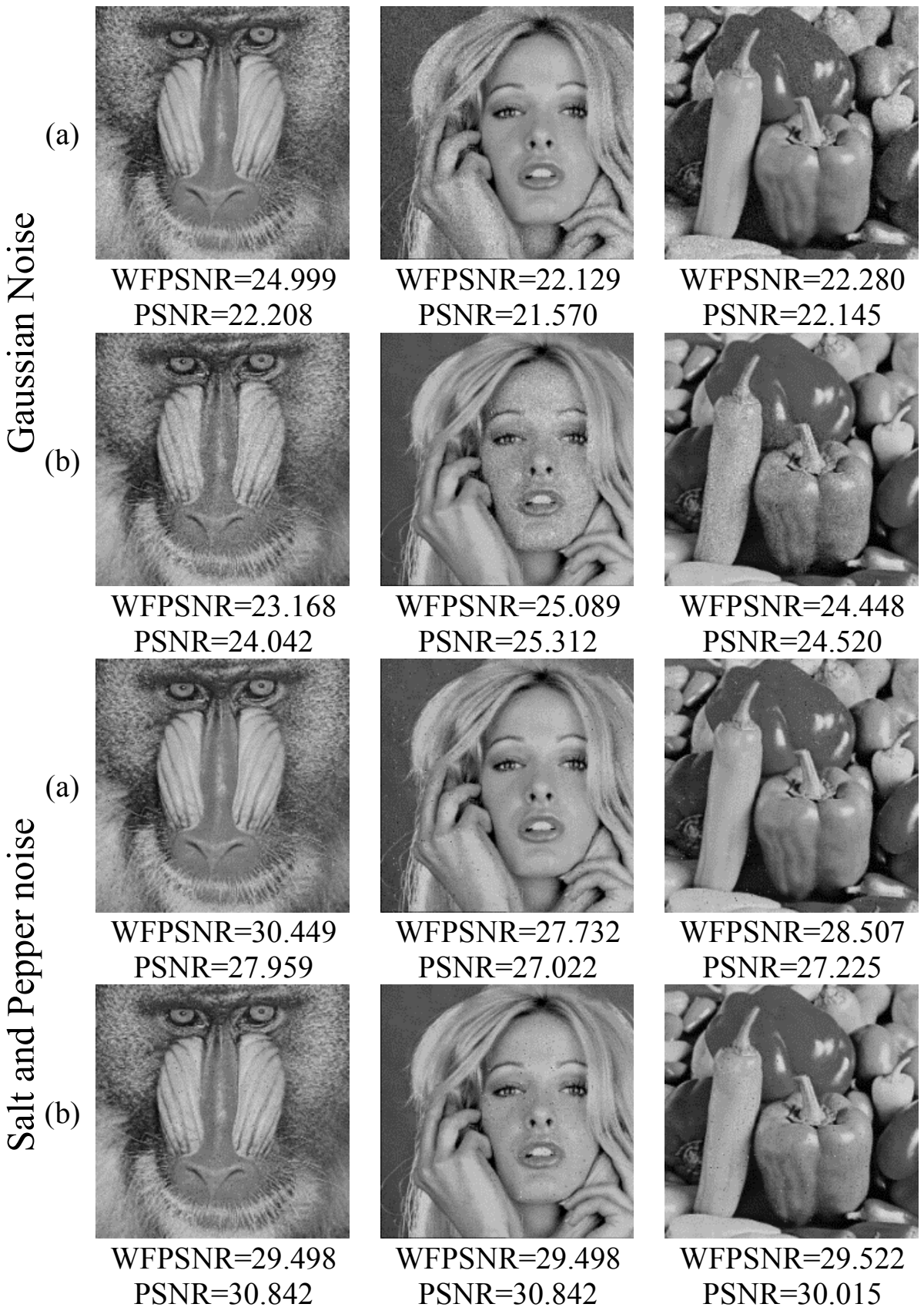

Figure 8. Comparison of standard and WFPSNR values after attacks on important and non-important areas from the user's point of view, (a) attacks on non-important areas, (b) attacks on important areas

## IV. Conclusion

In this paper, we proposed a weighted PSNR that is suitable for watermarking. This criterion considers parts of the image with a high degree of importance based on HVS. A combination of three features, saliency, edge concentration, and intensity, are fed into a fuzzy system to make a fuzzy map based on the human observer's requirements. By comparing the standard PSNR and proposed weighted PSNR, our method has significant results based on HVS. In our future works, we also can make content base PSNR that considers the important parts of the image based on human wants and after segmenting the image and finding these parts, give them higher weights in the final PSNR map. The proposed method can be applied to other watermarking methods [14-18].